# Electrical properties of thermal oxide scales on pure iron in liquid lead-bismuth eutectic

Jie Qiu[a], Junsoo Han[c], Ryan Schoell[d], Miroslav Popovic[a], Elmira Ghanbari[a], Djamel Kaoumi[d], John R Scully[c], Digby D Macdonald[a,*], Peter Hosemann[a,b,*]

[a] *Department of Nuclear Engineering, University of California at Berkeley, Berkeley, CA 94720, USA*

[b] *Lawrence Berkeley National Laboratory, Berkeley, CA, 94720, USA*

[c] *Center for Electrochemical Science and Engineering, Department of Materials Science and Engineering, University of Virginia, Charlottesville, VA 22904, USA*

[d] *Department of Nuclear Engineering, North Carolina State University, Raleigh, NC 27607, USA*

**Abstract**

The impedance behavior of pre-oxidized iron in liquid lead-bismuth eutectic (LBE) at 200 °C is studied using electrochemical impedance spectroscopy. The structures and resistance of oxide grown on iron oxidized in air at different temperatures and durations are compared. The results show that the resistance of the oxide film increases with increasing oxidizing temperature, due to the formation of a thicker scale and fewer defects. At the same temperature (600 °C), increasing the oxidation time can also reduce the defect concentration in the oxide film and improve the impedance of the oxide scale in LBE.

---

\* Corresponding authors:

macdonald@berkeley.edu (D.D. Macdonald); peterh@berkeley.edu (P. Hosemann).

# 1. Introduction

Lead bismuth eutectic (LBE) has been proposed as a candidate coolant for the Gen-IV fast reactor as well as both coolants and neutron spallation target in the accelerator-driven system (ADS), due to its favorable physical, nuclear and chemical properties [1-4]. LBE is characterized by low melting point, high boiling point, high thermal conductivity, low viscosity, good gamma shielding, and high neutron yield [4-6]. LBE-cooled reactors or concentrated solar power systems will operate at high temperatures, which translates into higher thermodynamic efficiency, while requiring a less robust pressure boundary of the coolant system because of the lower vapor pressure of the coolant [7]. However, LBE at high temperatures is a highly corrosive environment [8,9]. Materials corrosion in liquid LBE occurs primarily through dissolution of alloys components (i.e. Ni, Mn, Cr) into the liquid [8-10]. The dissolution rate depends on the the ratio of the surface area of alloy to the volume of the liquid LBE, the compostions of the alloy and the content of impurities such as oxygen in the liquid LBE [8]. Additionally, the LBE can also penetrate into the materials in the zones with a high defect density, such as along grain boudaries, and induce intergranular corrosion [11]. The structural and compositional changes due to the selective dissolution and intergranular corrosion could lead to material failures. Therefore, the corrosion of structural materials in LBE is a great challenge for the application of LBE as a reactor coolant.

An extensive amount of work on mitigating the corrosion of materials in LBE has been reported [12-17]. An effective method of mitigating corrosion is to form an protective oxide layer on the surface of structural materials by controlling the oxygen concentration in LBE [18-23]. For example, Benamati et al. [21,22] exposed the austenitic AISI 316L steels to oxygen saturated LBE and found that a thin oxide layer formed on the surface and protected the austenitic steel from corrosion attack for 5000 h at low temperature (573 and 673 K). Müller et al. [23] corroded austenitic AISI 316L steel in a flowing LBE (2 m/s) with oxygen concentration at $10^{-6}$ wt% and found that an intact oxide layer formed on the steel surface which provided a good protection against corrosion attack after exposure in LBE for up to 2000 h at 420 and 550 °C. However, for practical LBE environments, it is almost impossible to obtain an ideal, pore-free and stress-free protective layer throughout the varying environment. These oxide films in LBE depend strongly on the temperature, oxygen concentration and the compositions of the materials. Although the oxide film can protect steels in LBE when the temperature is lower than 773 K, Benamati et al. [21,22] also found that the oxide film is unstable at high temperature, and the austenitic AISI 316L suffered a severe corrosion attack and the dissolution of steel occurred at 823 K. Gorynin et al. [24] revealed that the no effective oxide film was formed on steel surface if the oxygen concentration is lower than $10^{-7}$ at.% in liquid lead, and the oxidation of material itself becomes a problem in maintaining an alloy in a high oxygen-containg LBE. Under different oxygen concentration, the alloys with higher Cr, Si or Al concentration can form more compact and stable oxide films, which provide better corrosion resistance in long-term, high temperature LBE experiments [9,25,26]. To improve the protective efficiency of oxide film in LBE, the effect of oxygen concentration, temperature and alloy composition on the inhibition efficiency and structure of oxide scales of the candidate materials have been the subject of considerable attention [18-26]. However, most of these research studies employed *ex-situ* characterization.

Electrochemical impedance spectroscopy (EIS) is a powerful, *in-situ*, and non-destructive technique and has been extensively used to study the thermal oxide and passive film formed on metal surfaces in aqueous systems [27-29]. Since liquid metal is electronically conductive, EIS could be used as a complimentary method to monitor the electrical properties and integrity of the oxide scales in liquid LBE on a real-time basis [30,31]. Lillard et al. [31] applied the EIS to measure the impedance of the oxide scales on the HT-9 and 316L steels in LBE and demonstrated that an impedance response develops if the oxidation time is sufficiently long at high temperature to produce a thick oxide scale. Bolind et al. [32] studied the impedance responses of the oxide scales on pre-oxidized 9Cr-1Mo steel in LBE and found that the ferritic/martensitic steels produce more conductive oxide scales than does the austenitic steel. Chen et al. [33] employed EIS to monitor the impedance behavior of the FeCrAl alloy in LBE at 550 °C for 3600 h and observed a general increase of the impedance during the real-time EIS measurement, indicating the validity of using EIS to monitor the real-time corrosion kinetics of alloys in LBE. Some work on characterizing the impedance responses of oxide scales in LBE has already been reported. However, information on the effect of the oxide scale structure on the impedance behavior in the liquid LBE is limited. It is known that most of the candidate structural materials in LBE environment are Fe-based steels [8,9], which forms the typical oxide scale consisting mostly of Fe-oxides. Therefore, using EIS to obtain an in-depth understanding the impedance behavior of the oxide layers grown on iron is of fundamental interest to determine the properties of the oxide layer *in-situ* and elucidate the possibility that the Fe-oxides will fail in service.

In this work, the electrical properties of the thermal oxide layers on iron in liquid LBE were evaluated by EIS. The impedance of pure iron oxidized in air at different temperatures and for different durations are characterized and compared. The effect of the oxide scale structure on the impedance behavior in the liquid LBE is discussed.

## 2. Experimental

A high pure iron (99.99%) plate used in this study was acquired from Alfa Aesar. This plate was cut into samples with a dimension of 30 mm × 4 mm × 1 mm. Prior to the oxidation, the strip samples were ground with SiC papers down to 1200 grit and then ultrasonically cleaned with ethanol and deionized water in sequence. Following this, all the samples were pre-oxidized in a furnace in the air, and the oxidizing conditions are shown in Table 1. After pre-oxidation, one end of a strip was abraded off and welded to a wire to form the working electrode (WE) for the electrochemical experiments.

A three-electrode setup system was used in this work, and the schematic diagram of the experimental set-up is shown in Fig. 1. In this study, 500 g of LBE (45.5 wt.% Pb-55.5 wt.% Bi) was placed into a 304L stainless steel crucible and melted, open to the air, at the experimental temperature, and then the prepared WE was dipped approximately 10 mm into the liquid LBE to begin the experiments. A tungsten wire with a diameter of 0.5 mm was used as the reference electrode (RE). The 304L stainless steel crucible was connected to the counter electrode (CE). The experimental temperature was $200 \pm 3$ $^{\circ}$C.

Electrochemical measurements were performed using a Gamry Electrochemical Measurement System (PC4) and were displayed by Gamry Echem Analyst. All the experiments were started after the samples had been exposed to the liquid LBE for 1 h to reach a steady-state and to ensure good electrical contact. EIS was carried out at the open circuit potential (OCP) by scanning the frequency from 100 kHz to 0.1 Hz, with a sinusoidal signal amplitude of 10 mV. All electrochemical experiments were conducted in triplicate. The cross-sectional morphology of the pre-oxidized samples before and after the electrochemical experiments were examined using a FEI Quanta 3D FEG scanning electron microscopy (SEM) equipped with an Oxford Energy Dispersive Spectroscopy (EDS). The chemical compositions and structures of the oxide scales on the thermal oxidized iron were characterized by a field emission FEI Talos F200X transmission electron microscopy (TEM). TEM thin foils were cut using a FEI Quanta 3D focused ion beam (FIB) instrument from oxide layer of the thermal-oxidized samples.

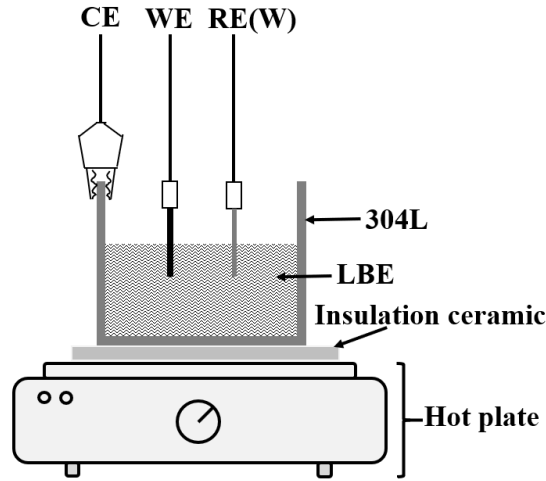

**Fig. 1**. The schematic diagram of the experimental set-up.

**Table 1:** The temperatures and times for oxidation of iron in this study.

| Samples | Temperature (°C) | Time (h) |
|---|---|---|
| 1# | 200 | 1 |
| 2# | 400 | 1 |
| 3# | 600 | 1 |
| 4# | 800 | 1 |
| 5# | 600 | 4 |
| 6# | 600 | 9 |
| 7# | 600 | 16 |
| 8# | 600 | 25 |

## 3. Results

### *3.1. Effect of oxidation temperature*

Fig. 2 shows the cross-sectional SEM images of the pure iron oxidized at different temperatures in air for 1 h. As expected, the thickness of oxide scale increases with increasing the oxidation temperature. There is no visible oxide film on iron oxidized at 200 °C for 1 h. With increasing the temperature to 400 °C, an oxide film with an average thickness of around 0.7 µm is formed on the iron, but some cracks are observed at some locations of the oxide film (Fig. 2(c)). For T > 400 °C, a compact homogeneous oxide scale forms on the surface. The thickness of the oxide scale at 600 °C and 800 °C are around 8.0 µm and 85.9 µm, respectively.

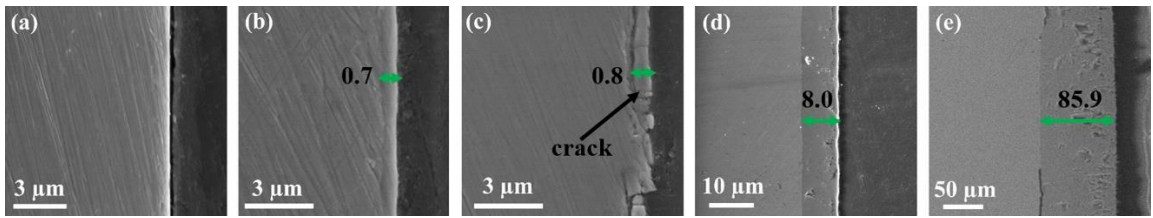

**Fig. 2.** Cross-sectional SEM images of iron oxidized at (a) 200 °C, (b) 400 °C, (c) 400 °C, (d) 600 °C and (e) 800 °C in air for 1 h.

To provide a better view of the effect of temperature, the oxide scales formed at different temperatures were further examined using TEM. Because the oxide formed at 200 °C is not visible, we have not characterized the sample, on which a very thin $Fe_2O_3$ may exist based on the literature [34]. Fig. 3 shows the TEM analyses of pure iron oxidized at 400 °C for 1 h. The bright field (BF) TEM image in Fig. 3(a) reveals that a compact oxide layer with small grain size formed on the iron. Selected area electron diffraction (SAED) pattern in Fig. 3.(c) reveals that the oxide is magnetite, which is consistant with the literature [35].

Fig. 4 shows the TEM analyses of pure iron oxidized at 600 °C for 1 h. It is found that the oxide scale includes a porous outer layer and compact inner layer. According to the bright field (BF) TEM image, the inner layer is a compact oxide with a grain size of around 0.3-2 μm. Based on the SAED in Fig. 4(d) and Fig. 4(e), the inner oxide formed on iron at 600 °C is a mixture of FeO and $Fe_3O_4$. As shown in Fig. 4(b), the outer layer is fine-grain oxide with a high density of pores. The SAED pattern in Fig. 4(f) reveals that the outer layer is magetite.

Fig. 5 and Fig. 6 show the TEM analyses of the outer layer and inner layer of the oxide scales formed on pure iron oxidized at 800 °C for 1 h, respectively. As shown in Fig. 5 (b), the inner oxide layer formed at 800 °C in a compact oxide with a larger grain size compared to the oxide that formed at 600 °C. The grain size of the oxides formed at 800 °C is from 0.5 μm to several μm. Based on the SAED pattern, the inner oxide scale corresponds to FeO, which is consistent with the literature reporting that the inner layer is wüstite [35], as shown in Fig. 5(d). Moreover, some magnetite was also observed in the inner oxide layer. As shown in the SAED pattern of Fig. 5(e), two sets of diffraction spots are visible along the $[\bar{1}12]$ zone axis. Besides the clear FeO diffraction pattern, there is a dimmed pattern, that was confirmed to be $Fe_3O_4$. It reveals that the inner oxide scale formed on iron at 800 °C comprised mainly FeO grains interspersed with very few $Fe_3O_4$ grains. Compared to the inner oxide layer, the outer layer has small grain size and a few faceted voids, especially within the first 2 μm surface area, as shown in Fig. 6(b). Although there are some voids in the outer layer, the oxide scale formed at 800 °C has fewer pores and larger grains than that formed at 600 °C. The SAED pattern in Fig. 6(d) and Fig. 6(f) show that the outer layer composed of $Fe_2O_3$ and $Fe_3O_4$. Based on these results, we conclude that increasing temperature increases the grain size and reduces the number of pores in the oxide scales.

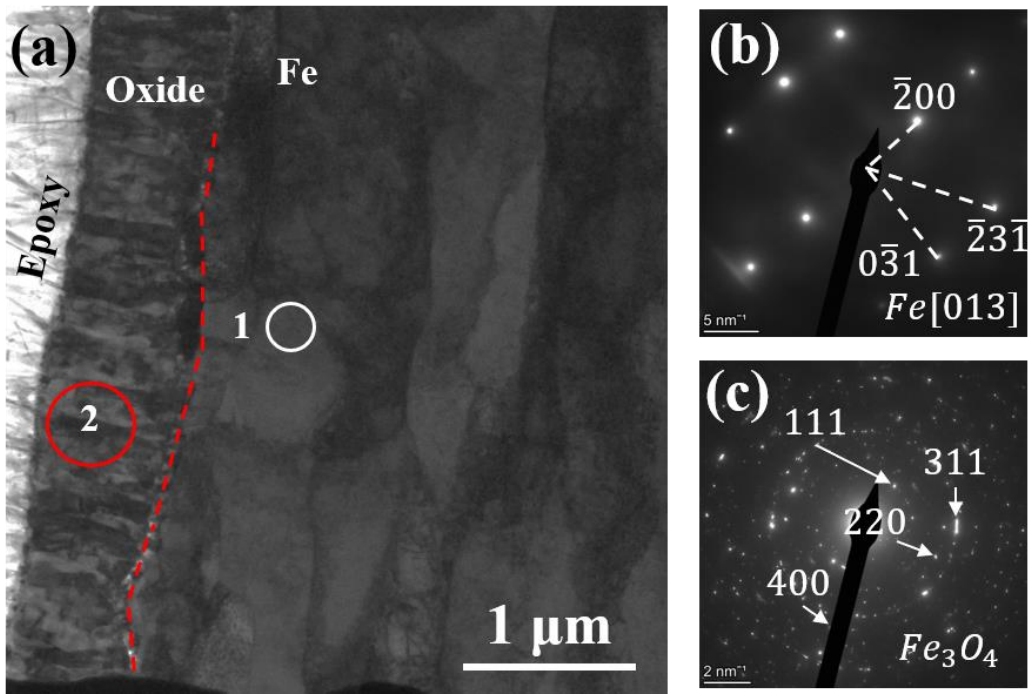

**Fig. 3.** (a) Low Magnification BF TEM Micrograph of pure iron oxidized at 400 °C for 1 h, (b) SAED of the Fe matrix shown as the circle 1 in (a), and (c) SAED of the oxide shown as the circle 2 in (a).

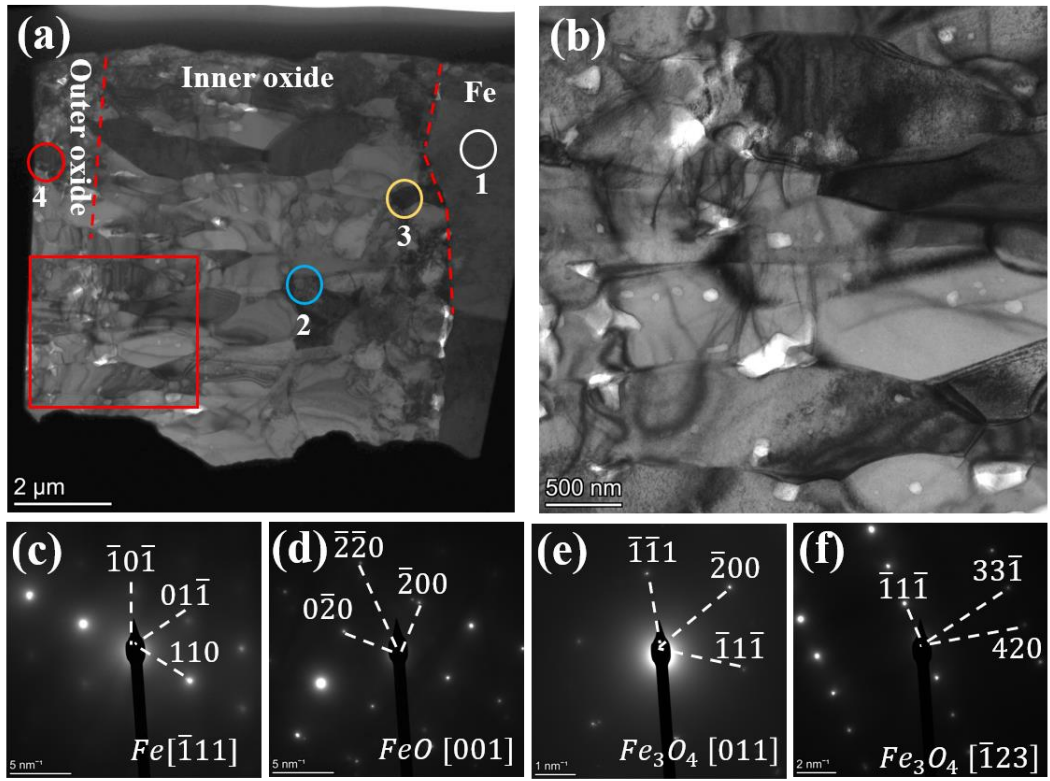

**Fig. 4.** (a) Low Magnification BF TEM Micrograph of pure iron oxidized at 600 ºC for 1 h, (b) Higher Magnification BF TEM Micrograph of outer oxide layer shown as red square in (a), (c) SAED of the Fe matrix shown as the circle 1 in (a), (d) SAED of the inner oxide layer as the circle 2 in (a), (e) SAED of the inner oxide layer as the circle 3 in (a), and (f) SAED of the outer oxide layer as the circle 4 in (a).

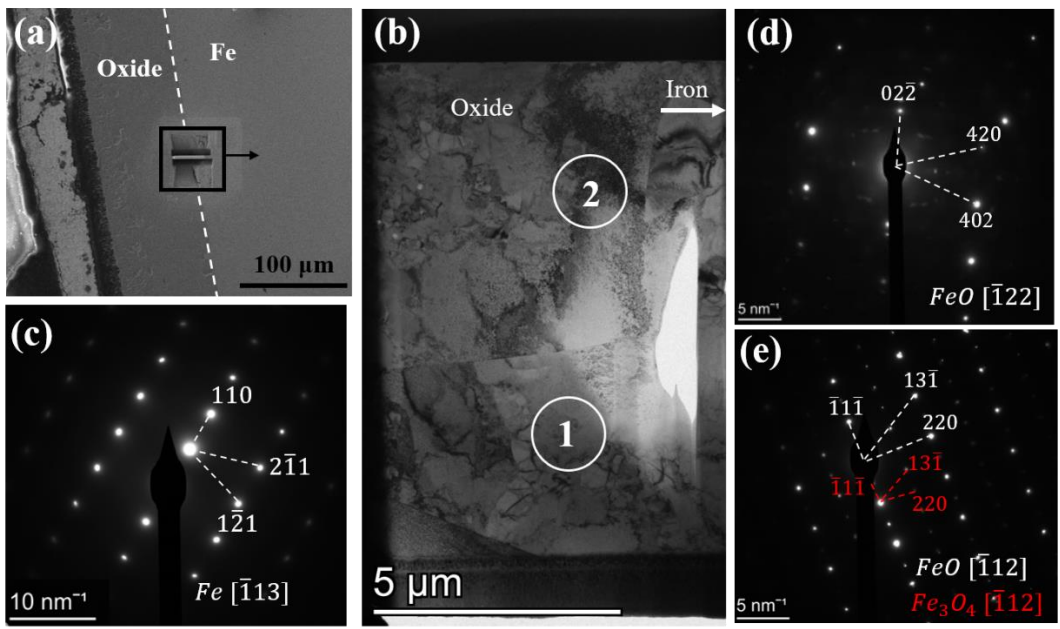

**Fig. 5.** (a) SEM image of TEM Liftout of pure iron oxidized at 800 ºC for 1 h, (b) Low Magnification BF TEM Micrograph of inner oxide layer shown as the black square in (a), (c) SAED of Fe matrix (location not shown), (d) SAED of inner oxide region shown as the white circle 1 in (b), and (e) SAED of inner oxide region shown as the white circle 2 in (b).

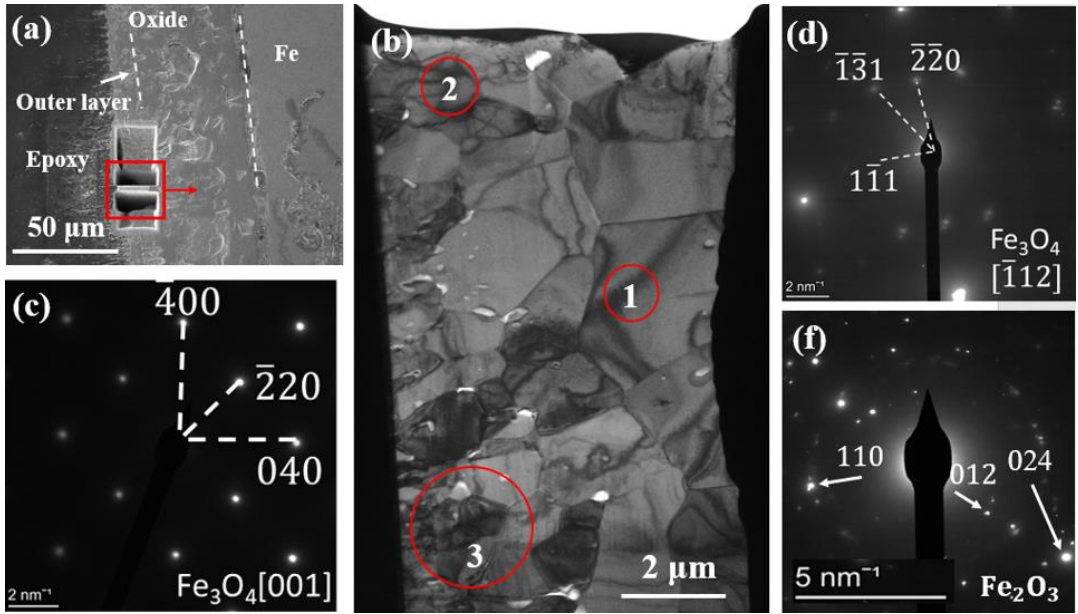

**Fig. 6.** (a) SEM image of TEM Liftout of pure iron oxidized at 800 °C for 1 h, (b) Low Magnification BF TEM Micrograph of outer oxide layer shown as red square in (a), (c) SAED of oxide region shown as red circle 1 in (b), (d) SAED of oxide region shown as red circle 2 in (b), and (e) SAED of oxide region shown as red circle 3 in (b).

To confirm the effect of temperature on the oxide scales, the pre-oxidized iron samples were immersed into the liquid LBE at 200 °C for the EIS measurements. Fig. 7 shows the Nyquist and Bode plots of the pre-oxidized iron in LBE. These data were recorded from the samples after 1 h immersion in liquid LBE to achieve a steady state. As shown in Fig. 7(a), the impedance modulus of the oxide film increases with increasing oxidation temperature. The phase angle of the impedance of the oxide scale formed at higher temperatures are more negative than those at lower temperatures, especially in the frequency range from 100 Hz to 100 kHz. The same features are also reflected in the Nyquist plots of Fig. 7(b), the loci corresponding to higher temperature have larger radii.

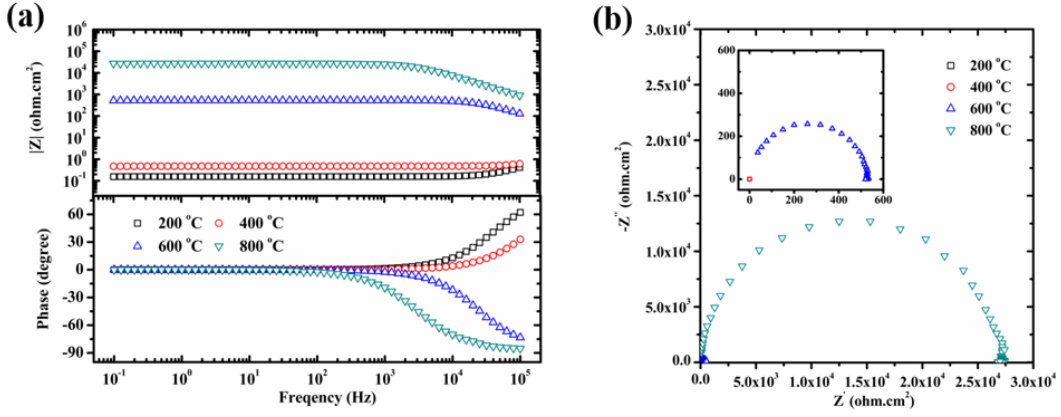

**Fig. 7.** (a) Bode and (b) Nyquist plots of pre-oxidized iron (Oxidized at different temperatures for 1 h) in liquid LBE (Note that there is only one data point the Nyquist plots for 200 °C and 400 °C, because the oxide film is so thin that it presents little impediment to the flow of electronic charge across the film, and hence is meaningless as far as the properties of the oxide are concerned).

As liquid LBE is an electronic conductor, there is no Helmholtz double layer at the sample/liquid metal interface. A Randles electrical equivalent circuit (EEC) successfully describes the EIS data on iron in liquid LBE [32] and is chosen in this work, as shown in Fig. 8(a). In this circuit, $R_{oxide}$ is the oxide scale resistance, $C_{oxide}$ is the oxide scale capacitance, $R_{LBE}$ is the resistance of liquid LBE. Taking into account the dispersion effect, due to variations in the oxide thickness, a constant phase element (CPE) was used instead of the pure oxide scale capacitor in this EEC analysis [32]. The impedance of CPE ($Z_{CPE}$) is expressed as follows:

$$Z_{CPE} = \frac{1}{Q_o(jw)^n} \quad (1)$$

where n is the CPE exponent, $w$ is the angular frequency and $Q_o$ is the CPE constant. According to the Hsu and Mansfeld [36], the effective capacitance is equal to:

$$C_{oxide} = \frac{(R_{oxide}Q_o)^{1/n}}{R_{oxide}} \qquad (2)$$

By fitting the EIS data using the Zview software, it is found that this EEC accounts for the EIS data very well. A comparison of the measured and fitted impedance is shown in Fig. 8(b).

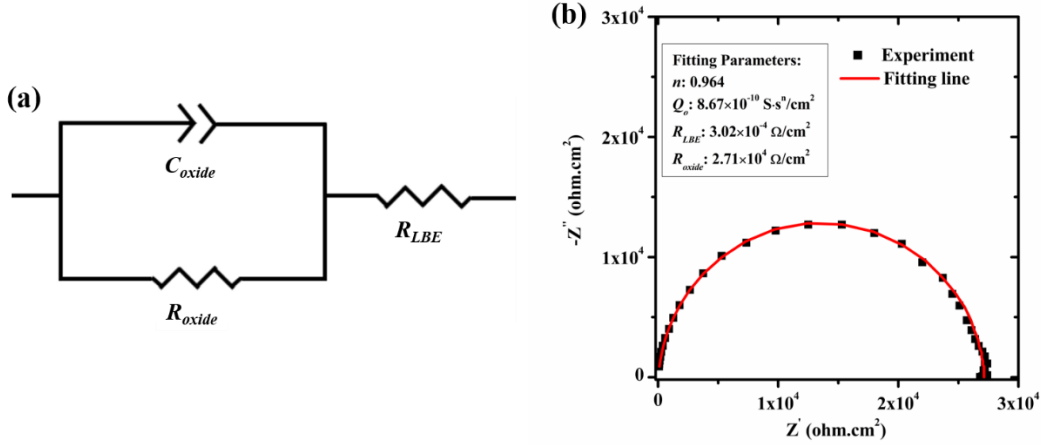

**Fig. 8.** (a) The Electrical Equivalent Circuit used to fit the impedance of the oxide film in liquid LBE, (b) Comparison of measured Nyquist plot of the pre-oxidized iron (800 °C for 1 h) in 200 °C LBE and the fitting line using the EEC in (a).

By fitting, a summary of the $R_{oxide}$ of the pre-oxidized iron at different temperatures is presented in Fig. 9(a). For comparison, the average thickness of the oxide film on pure iron from SEM imaging are also summarized in blue in Fig. 9(a). The thickness of the oxide film is the average value of seven different locations in the SEM image, including the thinnest and thickest locations. It is found that increasing oxidation temperature resulted in a thicker oxide, and therefore, an increases the impedance of the thermal-oxidized iron in liquid LBE. Moreover, when we divide the impedance by thickness, it is found that the per micrometer thickness resistance (oxide resistance divided by thickness) of oxide film on iron oxidized at different temperature is not constant, as shown in Fig. 9(b). The per micrometer thickness resistance of the oxide film formed at 800 °C was much higher than that formed at 600 °C, indicating that thickness, alone, is not the only factor controlling the oxide resistance.

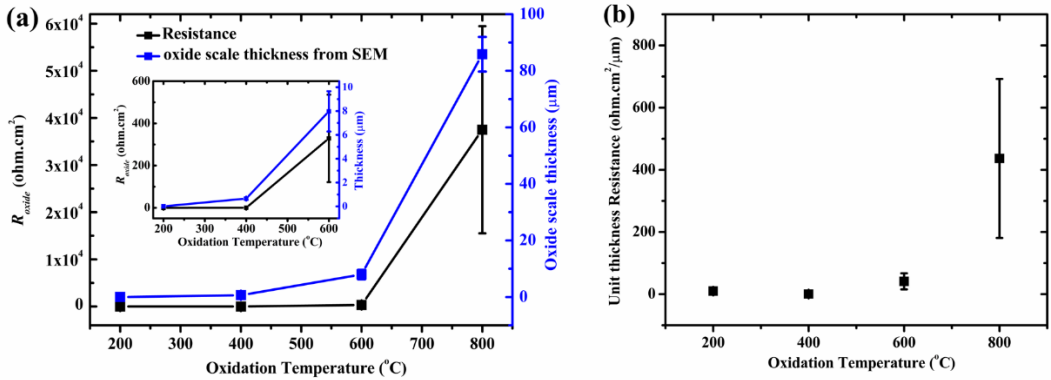

**Fig. 9.** (a) The resistance and thickness and (b) the unit thickness resistance of pre-oxidized iron as a function of oxidation temperature for an oxidation time of 1 h. The error bars were determined from triplicate experiments.

## *3.2. Effect of oxidation time*

Fig. 10 shows the EIS of iron pre-oxidized in air at 600 °C for different times. The impedance modulus of the oxide film on iron increases with increasing pre-oxidation time, and the phase angles of the oxide scale formed at longer time are more negative than those formed at shorter pre-oxidation time. The same features are also reflected in the Nyquist diagrams. It is seen that the loci corresponding to longer oxidation time has larger radii, which signifies higher corrosion resistance of the oxide film in LBE. The $R_{oxide}$ of the oxide scale on iron pre-oxidized at 600 °C for different times are obtained by fitting the EEC given in Fig. 8(a) on the experimental EIS data to confirm the effect of oxidation time. The results are summarized in Fig. 11(a) (black points, first y-axis). It is shown that the resistance of the oxide film increases exponentially with oxidation time.

To confirm the effect of oxidation time, the cross-sectional SEM images of the pre-oxidized iron at 600 °C for different oxidation times were examined, as shown in Fig. 12. A compact homogeneous oxide scale forms on iron, and the thickness of the oxide film increases upon extending the pre-oxidation time. Much has been written [35,37,38] on the kinetics of the thermal oxidation of pure iron. A commonly accepted mechanism is that the scale is found to grow initially at a constant rate (dx/dt = k), which gives rise to the linear growth law of $x = kt + x_0$, and then transforms into a parabolic relationship at longer oxidation times, although the oxidation behavior of iron at low temperature (< 700 °C) is complex and less consistent [35]. To confirm the oxidation behavior, the average thickness of the oxide film on pure iron was calculated according to the SEM images, and the results are summarized in Fig. 11(a) (blue points referred to the second y-axis). The oxide thickness is given in an average of seven different locations in the SEM image, including the thinnest and thickest. As shown in Fig. 11(a), the oxidation thickness of iron at 600 °C as a function of oxidation time can be fitted by the Tammann-type parabolic equation, which indicates that the oxidation corresponding to an ion diffusion mechanism [35]. Due to the relatively thick scales ($L > 3 - 4$ µm) formed at 600 °C, the transport of ions through the oxide scale is slow compared with any surface reaction and hence the oxidation rate is determined by diffusion mechanism [35]. To further confirm how the oxidation time affects the resistance of the oxide scale, the per micrometer thickness resistance of oxide film on pure iron oxidized at 600 °C for different times was calculated and is shown in Fig. 11(b). The plot reveals that the per micrometer thickness resistance of the oxide film formed at 600 °C increases with increasing the oxidation time rather than being constant, which demonstrates that thickness is not the only factor controlling the oxide resistance.

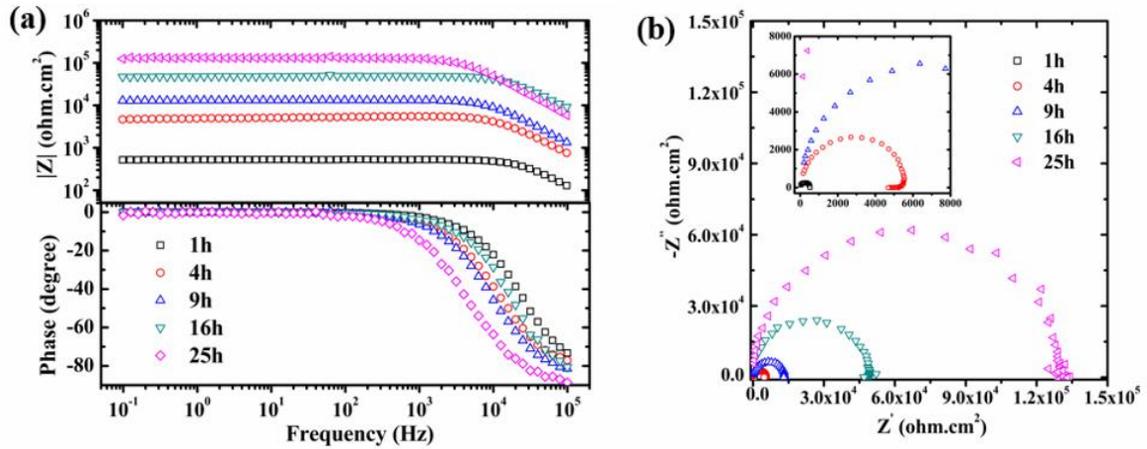

**Fig. 10.** (a) Bode and (b) Nyquist plots of oxide scales on iron in liquid LBE as a function of pre-oxidation time at 600 °C.

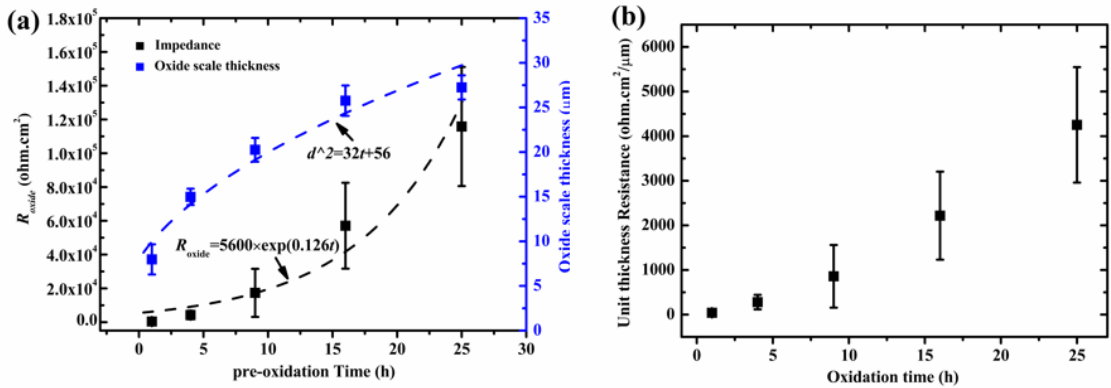

**Fig. 11.** (a) The resistance and thickness and (b) the unit thickness resistance of iron oxidized at 600 °C as a function of oxidation time. The resistance error bars were determined from triplicate experiments. Thickness was the average value of seven

different locations in SEM image, including the thinnest and thickest; the resistance is the average value of triplicate experiments.

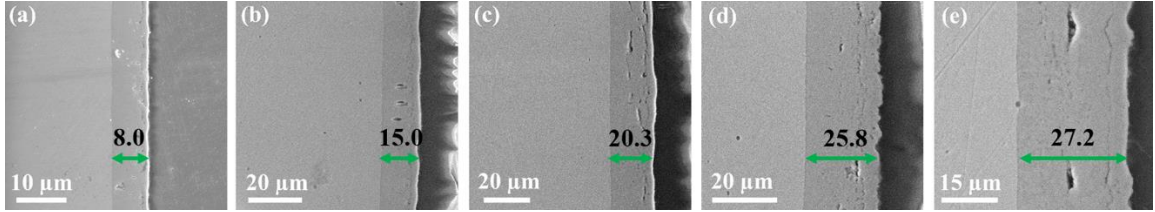

**Fig. 12.** Cross-sectional SEM images of iron pre-oxidized at 600 °C for (a) 1 h, (b) 4 h, (c) 9 h, (d) 16 h and (e) 25 h before the measurements of the impedance in LBE.

## 4. Discussion

Based on the above results, it is known that the impedance of oxide scale increases with increasing the oxidation temperature. At 200 °C, as shown in Fig. 2(a), there is no visible oxide film on iron oxidized at 200 °C for 1 h n implying a thickness << 100 nm perhaps with roughness, and therefore we can barely detect the impedance using EIS in LBE. Although a nanoscale-thick passive film formed at low temperatures protect materials from corrosion in aqueous solution, it does not appear impede current flow in liquid LBE. With increasing the temperature to 400 °C, the impedance is smaller than 1 Ω cm$^2$, as shown in Fig. 7. Based on the SEM images in Fig. 2(b) and Fig. 2(c), the oxide film formed at 400 °C is non-uniform and with some cracks. Thus, the oxide scale formed on iron at 400 °C is readily electrically short-circuited in LBE, resulting in the extremely low resistance ($|Z| < 1$ Ω·cm$^2$). This observation is consistent with other result from the litereature [39], which reported that the impedance of pre-oxidized 316L steel is negligible once cracks existed within the oxides. When the temperature is further increased to 600 ºC and 800 ºC, a compact, homogeneous, thick oxide scale forms on the sample surface, resulting in a higher impedance of the thermal-oxidized iron in liquid LBE. This results agree well with the previous study by Nam et al [40]. In that study, Nam also found that the area-normalized resistance will increase to 800 Ω cm$^2$ when thick oxides formed on the pure iron in LBE. Besides the thickness, the results reveal that the impedance is also related to the structure of the oxide scales. As shown in Fig. 9(b), the per micrometer thickness resistance of the oxide film formed at 600 ºC and 800 ºC is not constant. TEM results show that the oxide scale formed at 800 ºC has fewer pores and a larger grain size than that formed at 600 ºC (Fig. 4, Fig. 5 and Fig. 6). Thermally grown oxides on iron in air always include various $Fe^{2+}$ and $Fe^{3+}$ ionic defects compensated with n-type electronic defects [29]. These defects control transport through the oxide layers. Increasing temperature apparently reduces the defect density of the oxide scales, thereby increasing the impedance of thermal-oxidized iron in liquid LBE.

Similarly, increasing the oxidation time at a constant temperature can also improve the corrosion resistance of the oxide scale in LBE. As shown in Fig. 11(a), the $R_{oxide}$ of the oxide scale on iron pre-oxidized at 600 ºC increases with increasing oxidation time. However, the resistance of the oxide film increases exponentially with oxidation time, while the oxidation thickness as a function of oxidation time follows the parabolic law, suggesting that the resistance is not a linear function of the thickness (Fig. 11(a)). The observed behavior is readily accounted for by an increase in thickness coupled with a decrease in point defect density of the oxide scales with increasing pre-oxidation time. We known that the oxidation rate of iron follows the parabolic law, which is typically high initially and then gradually decreases with increasing oxidation time. The defect density generated during oxide scale growth decreases with a corresponding decrease in the oxidation rate [35]. Therefore, the per micrometer thickness resistance of the oxide scale increases with increasing the oxidation time at 600 °C (Fig. 11(b)).

To confirm this, the Mott-Schottky (M-S) analysis was performed to estimate the defect density in the oxide film. In light of the many issues in applying classical M-S theory to oxide scales [41], it is difficult to justify the quantitative accuracy of the M-S data unequivocally. However, it has been determined that the M-S analysis can be used qualitatively or even semi-quantitatively, as in this study, to examine how the defect concentration in the film changes with film formation time and hence thickness. According to M-S theory, the space charge capacitance ($C_{sc}$) should vary linearly with voltage according to [42]:

$$\frac{1}{C_{sc}^2} = \frac{\pm 2}{\varepsilon' \varepsilon_0 eN}(V - V_{fb} - \frac{kT}{e}) \tag{3}$$

where the positive sign designates an n-type semiconductor, and the negative sign indicates a p-type semiconductor. $\varepsilon'$ is the dielectric constant of the oxide film (assume to be 12.7 in this work [43]), $\varepsilon_0$ is the vacuum permittivity ($8.85 \times 10^{-14}$ F cm$^{-1}$), $e$ is the electronic charge ($1.60 \times 10^{-19}$ C), $N$ is the dopant (defect) density, $V$ is the applied potential, $V_{fb}$ is the flat-band potential of the film, $k$ is Boltzmann constant ($1.38 \times 10^{-23}$ J K$^{-1}$), and $T$ is the Kelvin temperature (K).

We found that it was very difficult to obtain M-S data in liquid LBE, and the M-S has been used successfully to estimate the defect density of oxide films on iron in borate buffer solution [29,44]. Therefore, we measured the M-S of the pre-oxidized iron in a 0.1 M $Na_2B_4O_7$ solution (pH = 9.3) at ambient temperature for this study. According to Equation (3), the defect density was calculated from the slope of the linear region in M-S profiles, and the defect density as a function of oxidation time are shown in Fig. 13. Within experimental uncertainty, it is found that the defect density decreases with increasing oxidation time corresponding to a concomitant increase in the scale resistance.

It should be pointed out that all the EIS experiments were performed in LBE open to the air at 200 ºC. Due to the air (21 vol.% $O_2$) in contact with the liquid LBE at low temperature (200 ºC), the oxygen content in LBE during the test was saturated, conditions under which the oxide film is thermodynamically stable and is not dissolved. It is known that the oxygen concentration and temperature of LBE influences the dissolution and oxidation behavior of materials. As reported [19], if the oxygen concentration in LBE is lower than $10^{-8}$ wt.%, the oxide scale might dissolve and cannot protect materials from corrosion in LBE at high temperature. Thus, in the future, high temperature and atmosphere-controlled corrosion experiments will be necessary to confirm the iron oxide corrosion resistance in the real operation conditions of liquid LBE.

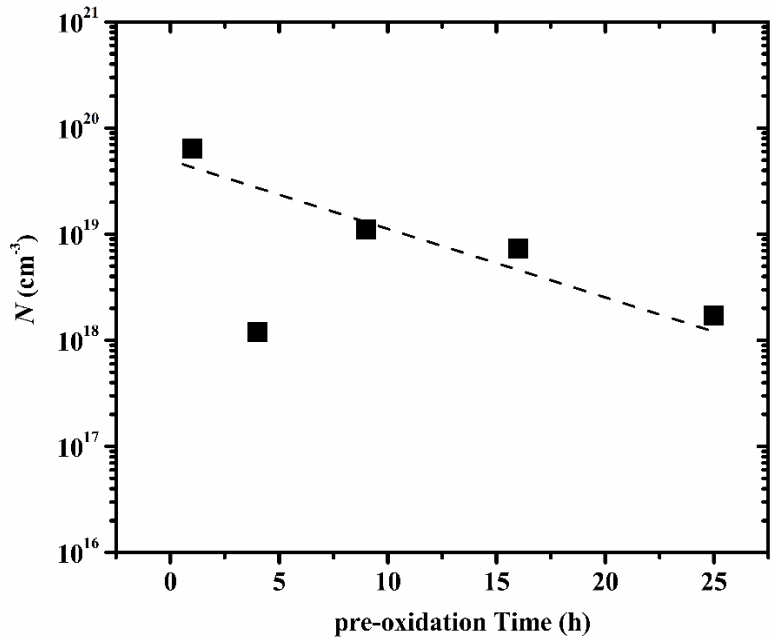

**Fig. 13.** The defect concentration of the oxide scale on pure iron oxidized at 600 °C as a function of oxidation time.

## 5. Summary and conclusions

The electrochemical behavior of pre-oxidized iron in liquid LBE at 200 °C was measured using EIS. The structures and resistance of the oxide grown on iron oxidized in air at different temperatures and durations were compared. The results reveal that increasing temperature could increase the grain size and reduce the number of pores in the oxide scales. With increasing oxidation temperature, the resistance of the oxide film increases, due to the formation of a thicker scale and fewer point defects over time. At 600 °C, the impedance magnitude of the oxide film on pure iron in liquid LBE increases with increasing pre-oxidation time, which is caused by the increasing thickness of the oxide film and decreasing defect concentration in the film at longer oxidation time. These results demonstrate that EIS is a promising method and could be used as a complimentary method for investigating the corrosion of materials in liquid metal systems.

## Acknowledgements

This work was supported as part of FUTURE (Fundamental Understanding of Transport Under Reactor Extremes), an Energy Frontier Research Center funded by the U.S. Department of Energy, Office of Science, Basic Energy Sciences. This work was performed in part at the Analytical Instrumentation Facility (AIF) at North Carolina State University, which is supported by the State of North Carolina and the National Science Foundation (award number ECCS-1542015). The AIF is a member of the North Carolina Research Triangle Nanotechnology Network (RTNN), a site in the National Nanotechnology Coordinated Infrastructure (NNCI).

## Data availability

The raw data required to reproduce these findings cannot be shared at this time as the data also forms part of an ongoing study.